\documentstyle[psfig,12pt]{article}
\title{Solar Magnetic Field Profiles and SuperKamiokande Data}
\author{\bf Jo\~{a}o Pulido \\
Centro de F\'{\i}sica das Interac\c{c}\~{o}es Fundamentais \\
Instituto Superior T\'{e}cnico \\
Av. Rovisco Pais, 1096 Lisboa Codex, Portugal}

\newcommand{\be}{\begin{equation}}
\newcommand{\ee}{\end{equation}}
\newcommand{\bea}{\begin{eqnarray}}
\newcommand{\eea}{\end{eqnarray}}
\begin{document}
\maketitle

\begin{abstract}
The distortion of the energy spectrum of recoil electrons in SuperKamiokande
resulting from several solar magnetic field profiles that are consistent with
data and standard solar models is investigated. The aim is to provide a test
of the general common features of these field profiles derived in a previous
work on the basis of the resonant spin flip flavour mechanism. It is found that
the distortion may be visible in the data, becoming possibly clearer when the energy
threshold succeeds in being reduced and distinct from the ones resulting from oscillations.  
\end{abstract}

\newpage

The possibility that neutrinos have a magnetic moment \cite {VVO} may provide
a unique insight on the inner solar magnetic field, with the resonant spin
flip flavour conversion of neutrinos \cite {LMA} taken as the main origin of the solar 
neutrino deficit. In general terms this deficit consists in the fact that too few neutrinos are being detected
\cite {Hom} - \cite{SAGE} as compared to the theoretical predictions \cite{BP95} -  
\cite{SDI}. The location of the resonance inside the Sun in a resonant process 
is uniquely fixed for a given flavour mass square difference by the neutrino energy, so that a high
suppression at a given energy range is interpreted in the resonant spin flip flavour as a high field intensity over a
corresponding spatial range. In this way it was shown \cite {Tom} that the results
from the four solar neutrino experiments \cite {Hom} - \cite {SAGE} are indicative of 
an average field intensity rising sharply by at least a factor 6 - 7 over a distance 
no longer than 7 - 10\% of the solar radius, decreasing then gradually towards the 
surface. Helioseismology suggests that such a sharp rise must lie around the upper layers
of the radiative zone and the bottom of the convective zone \cite {Parker}, with
the field reaching the order of $10^{5}$G at its maximum. The required order of
magnitude of the electron neutrino magnetic moment is from a few times $10^{-13}{\mu_B}$
to its laboratory upper bounds \cite {Allen}.

This magnetic field scenario is consistent with but obviously  in no way implied
by the data on solar neutrinos. Its verification can only be provided by the second
generation high precision experiments like SuperKamiokande \cite {SK} and SNO
\cite {SNO}. 

In this paper we investigate the distortion of the recoil electron energy spectrum in
the SuperKamiokande experiment associated with the general magnetic field profile
described above. The high statistics of SuperKamiokande, together with the lowering
of their recoil electron energy spectrum will hopefully make it possible, if this scenario  
is realistic, to trace and identify its characteristic distortion.

The starting point is the true event rate in SuperKamiokande which we will denote by
$S^{'}(T^{'})$ and is given by \cite {PM}
\be
S^{'}(T^{'})=\int_{E_{\nu_m}}^{E_{\nu_M}}dE_{\nu}\left(P(E_{\nu})\frac{d^2{\sigma_W}}
{dT^{'}dE_{\nu}}+\frac{d^2{\sigma_{-EM}}}{dT^{'}dE_{\nu}}\right)f(E_{\nu})
\ee
 
Here $T^{'}$ is the true recoil electron kinetic energy, $f(E_{\nu})$ is the energy
distribution \footnote{The uncertainties in the solar magnetic field do not justify
replacing the more recent standard neutrino energy spectrum from $^8B$ decay \cite {Bahcall96}
for the one used throughout this paper \cite {BU88}.} of $^8B$ neutrinos \cite {BU88}
and $P(E_{\nu})$ is the survival probability for neutrinos with energy $E_{\nu}$. The
lower and upper integration limits in eq.(1) are determined respectively by the 
kinematical inequality
\be
E_{\nu}{\geq}\frac{T^{'}+\sqrt{{T^{'}}^2+2m_eT^{'}}}{2}
\ee  

and the maximum $^8B$ neutrino energy \cite {BU88}
\be
E_{\nu_M}=15MeV.
\ee

The weak and electromagnetic spin flip parts of the ${\nu_e}e^{-}\rightarrow{\nu_e}e^{-}$
cross section are given by the expressions \cite{PM}
\be
\frac{d^2\sigma_W}{dT^{'}dE_{\nu}}=\frac{G_{F}m_{e}}{2\pi}\left((g_V+g_A)^2+(g_V-g_A)^2
(1-\frac{T^{'}}{E_\nu})^2-(g_V^2-g_A^2)\frac{m_eT^{'}}{{E_\nu}^2}\right)
\ee
\be
\frac{d^2\sigma_{-EM}}{dT^{'}dE_\nu}=f_\nu^2\frac{\pi\alpha^2}{m_e^2}\left(\frac{1}{T^{'}}-
\frac{1}{E_\nu}\right)
\ee
in standard notation and with $f_\nu$ being the neutrino magnetic moment in Bohr magnetons.
We assume vanishing mean square radius for the neutrino, so that the contributions of
the spin non-flip and interference cross sections are absent, and take for $\nu=\nu_e$
\be
g_V=\frac{1}{2}+2sin^2\theta_W~~~,~~~g_A=\frac{1}{2}
\ee
with $sin^2\theta_W=0.23$.

The true (physical) event rate $S^{'}(T^{'})$ given by (1) is in fact smeared by the
energy resolution function of the detector $R(T^{'},T)$ \cite{SK} and the measured event 
rate is instead
\be
S(T)=\int_{0}^{T^{'}_M}S^{'}(T^{'})R(T^{'},T)dT^{'}
\ee

where T is the measured recoil kinetic energy,
\be
T^{'}_M=\frac{2E^{2}_{\nu_M}}{2E_{\nu_M}+m_e}
\ee
(from inequality (2)) and \cite{SK}
\be
R(T^{'},T)=\frac{1}{\Delta_{T^{'}}\sqrt{2\pi}}exp\left(-\frac{(T^{'}-T)^2}{2\Delta^2_{T^{'}}}
\right).
\ee

In equation (9) the parameter $\Delta_{T^{'}}$ denotes the energy dependent $1\sigma$
width of the resolution function,
\be
\Delta_{T^{'}}=\Delta_{10}\sqrt{\frac{T^{'}}{10MeV}}.
\ee
Currently for SuperKamiokande $\Delta_{10}=1.5MeV$.

For the survival probability in equation (1), $P(E_\nu)$, we will use the Landau
Zener approximation whereby \cite{PhysRep}
\be
P(E_\nu)=P_{LZ}(E_\nu)=exp~(-\pi\frac{2{\mu^2}_{\nu}B^2}{\frac{\Delta^2}{2E_\nu}}0.09R_S)
\ee
with $B_{res}$ denoting the magnetic field at the resonance (critical) point,
\be
x_{res}=\frac{r}{R_S}=0.09~log~\frac{\frac{5}{3\sqrt{2}}2.11\times10^{-11}eV}
{\frac{\Delta^2}{2E_\nu}}~,
\ee
$\Delta^2$ the neutrino flavour mass square difference, $\mu_\nu$ the neutrino 
magnetic moment and $R_S$ the solar radius. In the present investigation we
assume a vanishing vacuum mixing angle, so we will be solely analysing the
joint effect of the neutrino magnetic moment and solar magnetic field.

The values of the event rate $S(T)$ given by expression (7) with $S^{'}(T^{'})$
and $R(T^{'},T)$ defined by (1) and (9) respectively will now be confronted with 
the corresponding event rate $S_{st}(T)$ for standard neutrinos $(\mu_\nu=f_\nu=0)$
for which the survival probability is obviously unity. We have
\be
S_{st}(T)=\int_0^{T^{'}_M}S^{'}_{st}(T^{'})R(T^{'},T)dT^{'}
\ee
with
\be
S^{'}_{st}(T^{'})=\int_{E_{\nu_m}}^{E_{\nu_M}}dE_{\nu}\frac{d^2\sigma_W}{dT^{'}dE_\nu}
f(E_\nu).
\ee

The ratio between equations (7) and (13) provides a measure of the deviation of the
recoil energy spectrum relative to the corresponding spectrum for standard neutrinos.
For high enough experimental sensitivity and sufficiently low energy threshold, as
will be seen, this deviation will become apparent, thus constituting a signature of
the suppression process occuring in the Sun. Another way of expressing this signature
is through the distortion of the electron energy spectrum. To this end, choosing to
normalize the above ratio between (7) and (13) to its value at T = 8 MeV, we will express
the distortion as
\be
D(T)=\frac{S(T)}{S_{st}(T)}\times\frac{S_{st}(8)}{S(8)}~.
\ee

We consider the general field profile described in the introduction with all its
possible variants, together with the corresponding solution ranges
in terms of $\mu_\nu$ and $\Delta^2$, compatible with solar neutrino data 
\cite{Hom} - \cite{SAGE} and standard solar models \cite{BP95}, \cite{TCL} -
\cite{SDI} (see table I). 
All these fields satisfy the general feature of having a sharp rise by
nearly an order of magnitude or more across the upper radiation zone and the bottom
of the convection zone, decreasing smoothly towards the surface with an upward facing
concavity or a linear decrease. (A downward facing concavity must be
excluded \cite{Tom}). The first five were taken directly from ref. \cite{Tom} and
the last one was added in order to account for the possibility of a $3\times10^5G$
field at the bottom of the convective zone \cite{Parker}.

It should be emphasized that the lowering of the energy threshold in the experiment
is essential in order to "magnify" the distortion effect for a given solar field 
distribution. In fact, for all field distributions, the survival probability
decreases rapidly with decreasing $E_\nu$ (see fig. 1 for a typical example). By
lowering the recoil electron  energy threshold $E_{e_{min}}$ ($T_{min}=E_{e_{min}}-m_e$),
the lower integration limit in equation (1)
decreases, so the smaller probabilities become more predominant in the integration,
providing a stronger reduction effect in $S(T)$ with respect to $S_{st}(T)$ at 
lower T.

The distortion of the recoil electron energy spectrum (eq. (15)) for the six field
profiles considered in table I is shown in figs. (2) through (7). The range of 
solutions in terms of $\mu_\nu$ and $\Delta^2$ is limited by the values denoted
(a), (b) in each case, the solid lines corresponding to cases (a) and the dashed
lines to cases (b). Hence the range of possible distortions for each field profile
is limited by the two curves. The "magnification" effect in the distortion for a
lowering T is apparent from these figures. 

As far as time variations of neutrino flux with solar activity are
concerned, no other suppression mechanism apart from the magnetic moment one can 
reflect them, since solar activity is correlated with magnetic field intensity.
It has not become clear as yet whether Homestake \cite{Hom}, the only
experiment claiming such time variations in its data so far, shows any evidence of
them. These data are consistent with no anticorrelation with the 11 year solar
cycle \cite{GW}, \cite{Haubold}, but other periodic time dependence of the data
may exist \cite{Haubold}. Only the second generation of experiments, namely
SuperKamiokande \cite{SK} may provide a clarification. The highest
energy $^8B$ neutrinos ($E_{\nu}{\geq}8MeV$), which are hardly suppressed (see fig. 
1) and therefore whose resonances are strongly non-adiabatic, cannot show any sign
of anticorrelation with solar activity and magnetic field. Once again it is only in 
the low $E_\nu$ sector and therefore in the lowest electron energy T sector of the
data that this effect may appear. A change in the average magnetic field by a factor
of approximately 3, as in the cases (5) and (6) of table I (see figs. 6 and 7
respectively), clearly shows up in the distortion, with the average larger field
(6) of the same shape as (5) corresponding to a larger distortion. So the spectrum
distortion of the recoil electron kinetic energy is correlated with the solar 
activity: the more intense the solar activity, the larger the distortion.
Its signature is also characteristically different from the one
originated by the vacuum oscillation, the large and the small mixing angle solutions
\cite{JNBSLAC}, \cite{KS}.    

To conclude, all possible solar magnetic field profiles that are compatible with
the data from the first generation of solar neutrino experiments provide a distortion
of the kinetic energy spectrum of recoil electrons that appears to be detectable
and distinguishable in SuperKamiokande and inceases with solar activity. To this end, it will be of primordial
importance to decrease as much as possible the energy threshold of the experiment.

{\vspace{10mm}}

{\newpage}
\begin{center}
\begin{tabular}{ccc} \hline\hline
B (Gauss)                                           &  x                           & solution \\
                                                    &                              & (a)~~~~~~~~~~~(b) \\ \hline 
0                                                   &  $0{\leq}x{\leq}0.7$         & $3.1\times10^{-12}\mu_B{\leq}
\mu_\nu{\leq}3.8\times10^{-12}\mu_B$ \\ 
$2\times10^6(x-0.7)$                                &  $0.7{\leq}x{\leq}0.75$      & $6.33\times10^{-9}eV^2{\leq}
\Delta^2{\leq}6.56\times10^{-9}eV^2$              \\ 
$10^5$                                              &  $0.75{\leq}x{\leq}0.8$      &  \\  
$10^5-4.9\times10^5(x-0.8)$                         &  $0.8{\leq}x{\leq}1$         &  \\ \hline  
$9.4\times10^4$                                     &  $x{\leq}0.645$              & $\mu_\nu=10^{-12}\mu_B$ \\
$9.32\times10^6(x-0.645)+9.4\times10^4$             &  $0.645{\leq}x{\leq}0.71$    & $1.5\times10^{-8}eV^2{\leq}
\Delta^2{\leq}1.91\times10^{-8}eV^2$               \\
$-3.4\times10^6(x-0.71)+7\times10^5$                &  $0.71{\leq}x{\leq}0.91$     &  \\
$-2\times10^5(x-0.91)+2\times10^4$                  &  $0.91{\leq}x{\leq}1$        &  \\ \hline
0                                                   &  $x<0.71$                    & $3.4\times10^{-11}\mu_B{\leq}
\mu_\nu{\leq}1.3\times10^{-10}\mu_B $ \\  
                                                    &                              &  \\
$\frac{3.048\times10^4}{cosh[20(x-0.71)]}$          &  $0.71{\leq}x{\leq}1$        & $1.6\times10^{-8}eV^2{\geq}
\Delta^2{\geq}7.3\times10^{-9}eV^2$  \\ \hline 
$2.16\times10^3$                                    &  $x{\leq}0.7105$             & $4.1\times10^{-12}\mu_B{\leq}
\mu_\nu{\leq}6.15\times10^{-12}\mu_B$  \\
$8.7\times10^4\left[1-\left(\frac{x-0.75}{0.04}\right)^2\right]$ &  $0.7105{\leq}x{\leq}0.7483$ & $6.5\times10^{-9}eV^2
{\leq}\Delta^2{\leq}6.9\times10^{-9}eV^2$\\ 
$10^5[1-3.4412(x-0.71)]$                            &  $0.7483{\leq}x{\leq}1$      & \\ \hline
$2.16\times10^3$                                    &  $x{\leq}0.7105$             & $7.2\times10^{-12}\mu_B{\leq}
\mu_\nu{\leq}2.1\times10^{-11}\mu_B$\\
$8.7\times10^4\left[1-(\frac{x-0.75}{0.04})^2\right]$ &  $0.7105{\leq}x{\leq}0.7483$ & $1.3\times10^{-8}eV^2{\geq}
\Delta^2{\geq}7.3\times10^{-9}eV^2$ \\
                                                    &                              & \\
$\frac{8.684\times10^4}{cosh[20(x-0.7483)]}$        &  $0.7483{\leq}x{\leq}1$      & \\ \hline  
$1.5\times10^3$                                     &  $x{\leq}0.7101$             & $2.1\times10^{-12}\mu_B{\leq}
\mu_\nu{\leq}7.3\times10^{-12}\mu_B$ \\
$3\times10^5\left[1-\left(\frac{x-0.75}{0.04}\right)^2\right]$   & $0.7101{\leq}x{\leq}0.7483$ & $1.3\times10^{-8}eV^2
{\geq}\Delta^2{\geq}7.2\times10^{-9}eV^2$ \\
                                                    &                              & \\
$\frac{2.9946\times10^5}{cosh[20(x-0.7483)]}$       &  $0.7483{\leq}x{\leq}1$      & \\ \hline  
\end{tabular}
\end{center}

Table I. Solar magnetic field profiles and corresponding solutions in 
terms of $\mu_\nu$ and $\Delta^2$ ranges consistent with data \cite{Hom} - \cite{SAGE} and 
standard solar models \cite{BP95},  \cite{TCL} - \cite{SDI}. The first five field profiles were proposed in
ref. \cite{Tom} and the last one in the present work using the same data and procedure. 
Notice that the fourth and fifth only
differ in the way they decrease along the convective zone while the last is the previous
one beyond the bottom of the convective zone multiplied by a scale factor 3.4.

\newpage

\begin{figure}
\begin{picture}(18,17)
\put(0,1){\psfig{figure=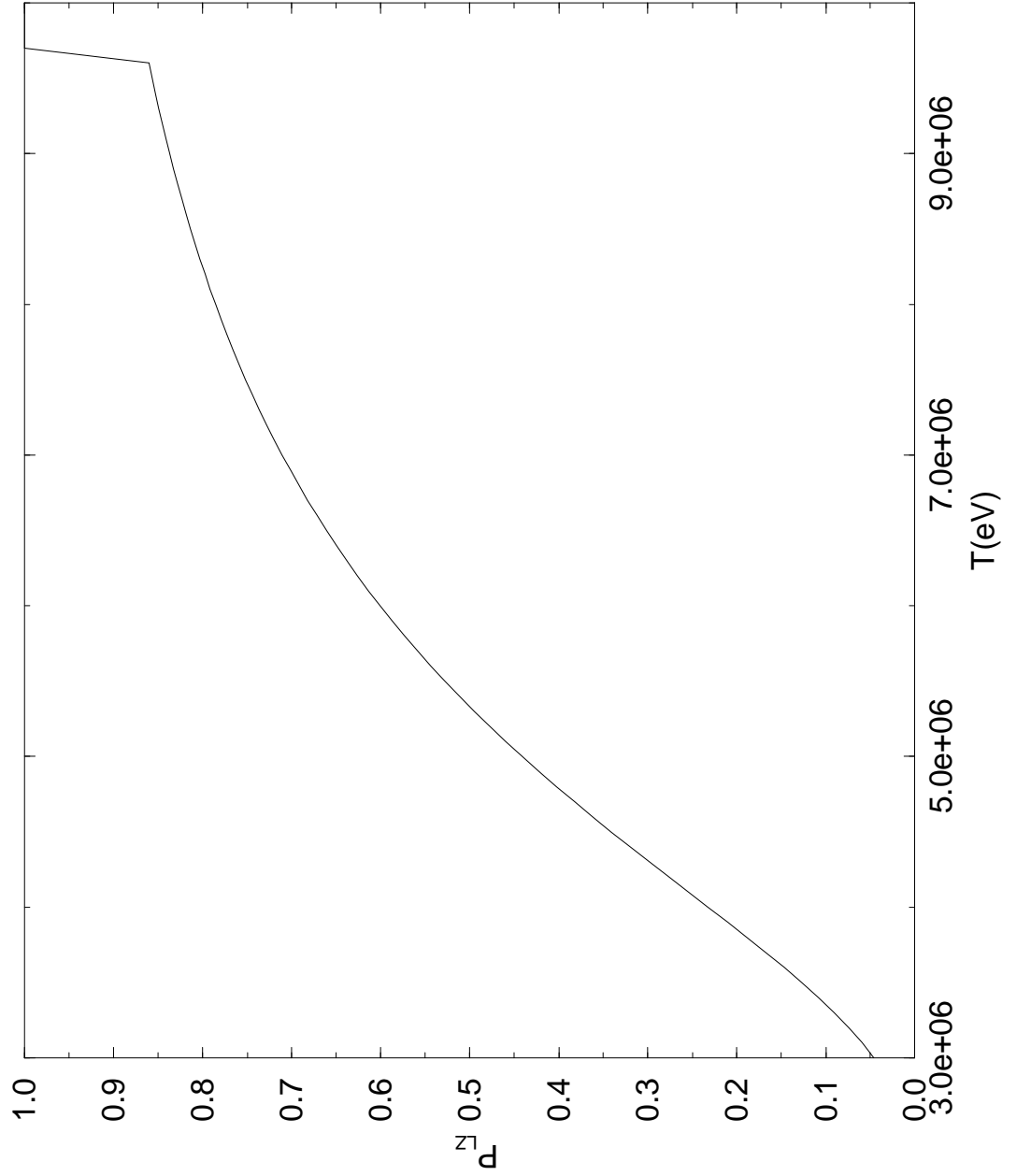,width=17cm}}
\end{picture}
\caption{The Landau-Zener probability as a function of neutrino energy $E_\nu$
for the last of the field profiles in table I at the upper end solution (case (b)):
$\mu_\nu=7.3\times10^{-12}\mu_B$, $\Delta^2=7.2\times10^{-9}eV^2$. The discontinuity in the 
derivative at the upper right is related to the sudden drop to zero of
the field at the Sun's surface. Units are in eV.}
\end{figure}

\begin{figure}
\begin{picture}(18,20)
\put(1,2){\psfig{figure=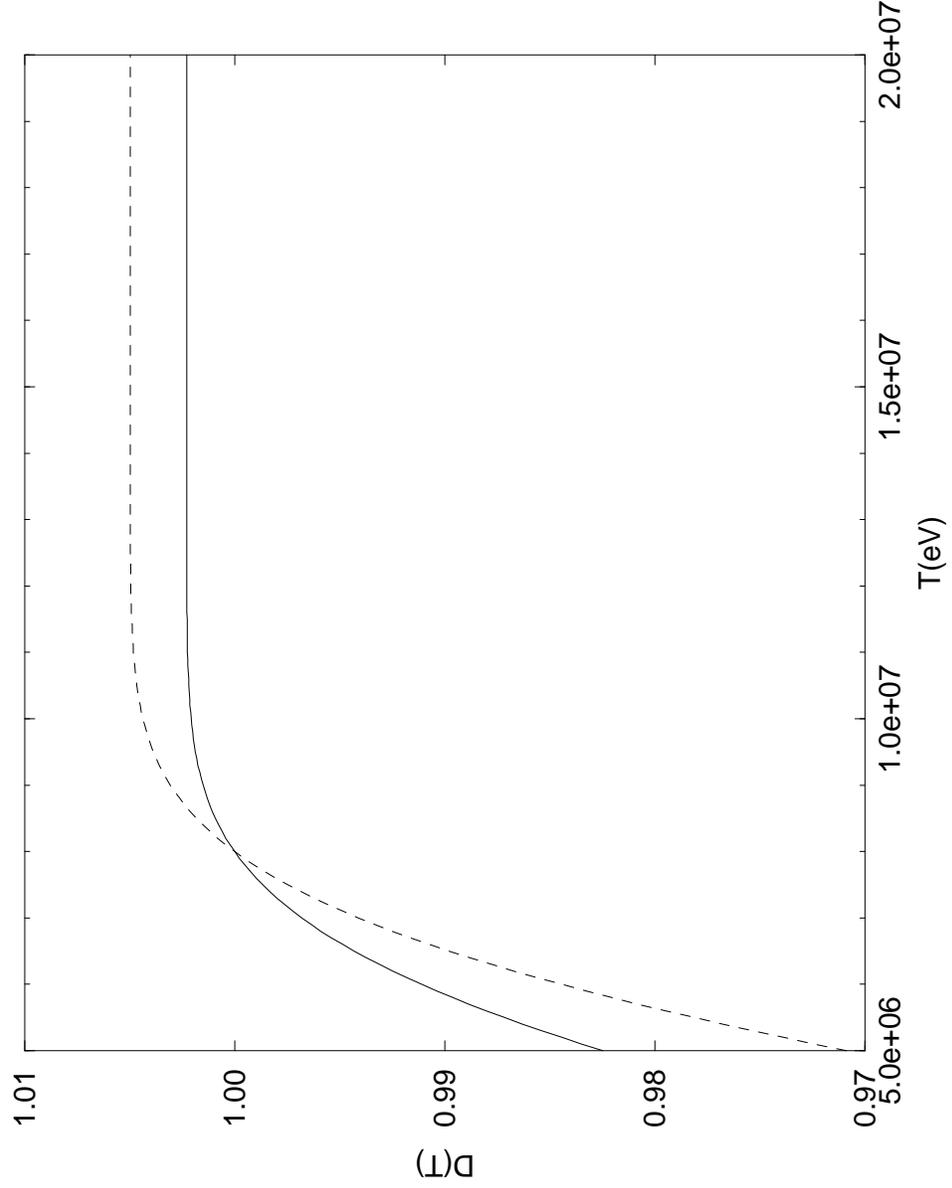,width=15cm}}
\end{picture}
\caption{The distortion of the recoil electron kinetic energy spectrum D(T) (eq. (15))
corresponding to the first of the field profiles in table I normalized to its value
at $T=8\times10^6eV$. The two ends of the solution range denoted (a) and (b) in table I 
correspond to the solid and dashed lines respectively. Units are in eV.}
\end{figure}

\begin{figure}
\begin{picture}(18,20)
\put(1,2){\psfig{figure=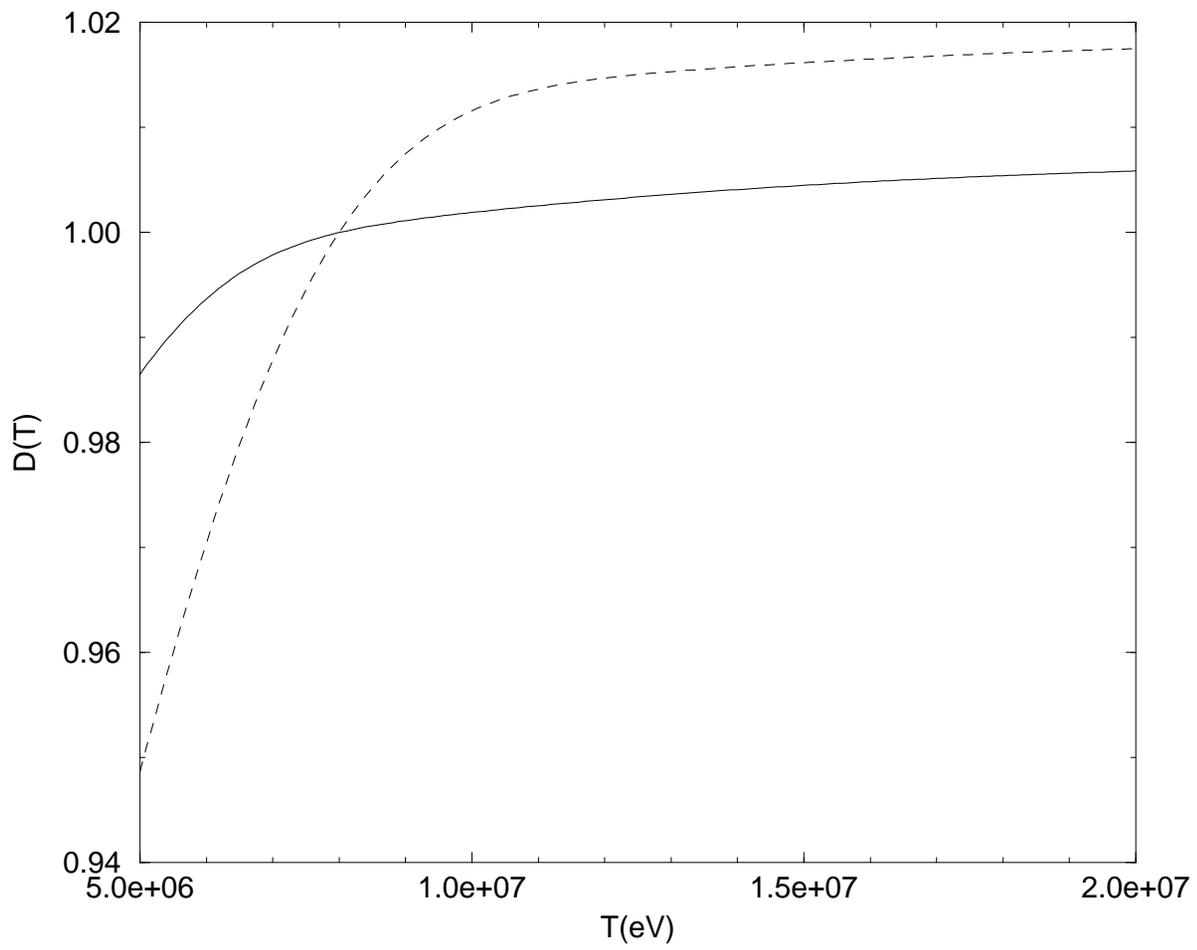,width=15cm}}
\end{picture}
\caption{Same as fig. 2 for the second field profile in table I.}
\end{figure}

\begin{figure}
\begin{picture}(18,20)
\put(1,2){\psfig{figure=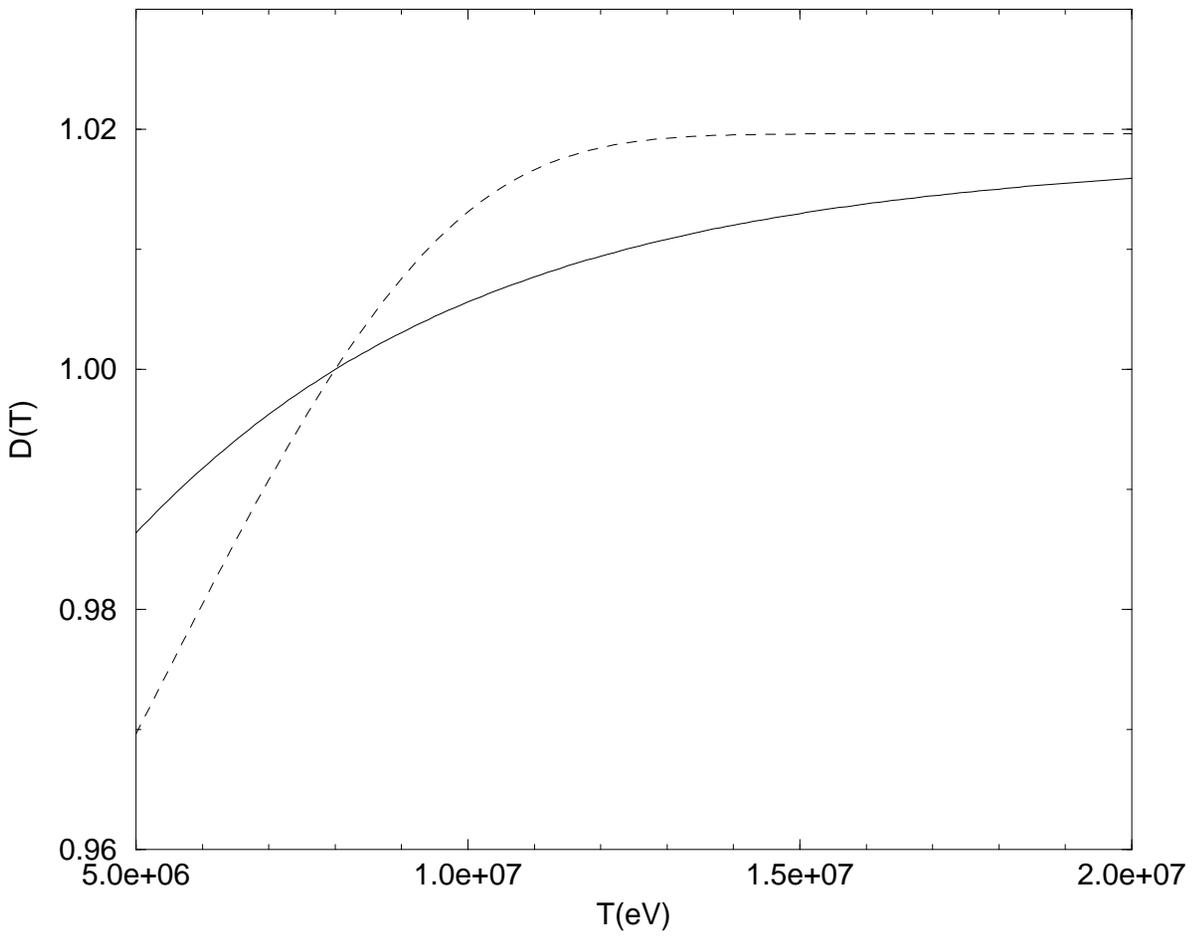,width=15cm}}
\end{picture}
\caption{Same as fig. 3 for the third field profile in table I.}
\end{figure}

\begin{figure}
\begin{picture}(18,20)
\put(1,2){\psfig{figure=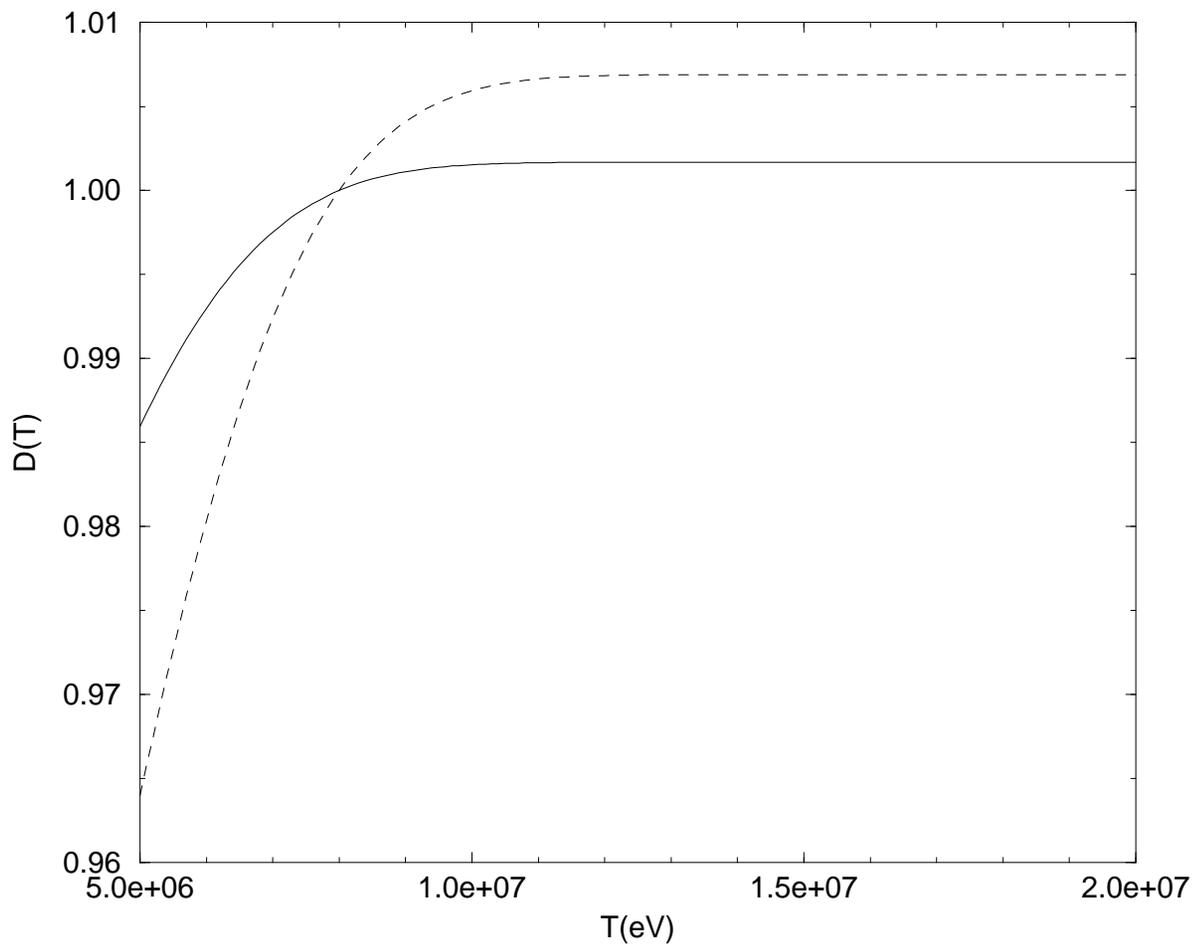,width=15cm}}
\end{picture}
\caption{Same as fig. 4 for the fourth field profile in table I.}
\end{figure}

\begin{figure}
\begin{picture}(18,20)
\put(1,1){\psfig{figure=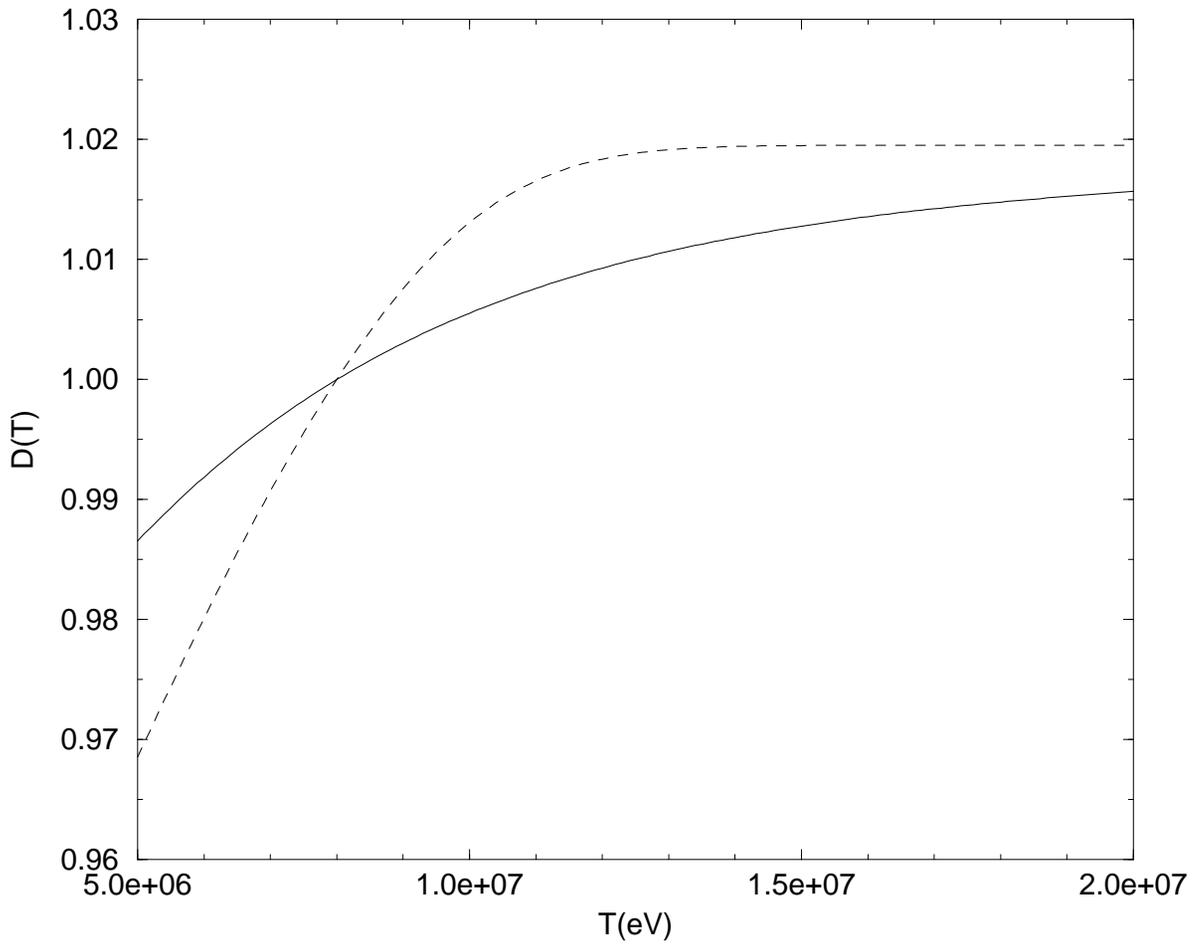,width=15cm}}
\end{picture}
\caption{Same as fig. 5 for the fifth field profile in table I.}
\end{figure}

\begin{figure}
\begin{picture}(18,20)
\put(1,2){\psfig{figure=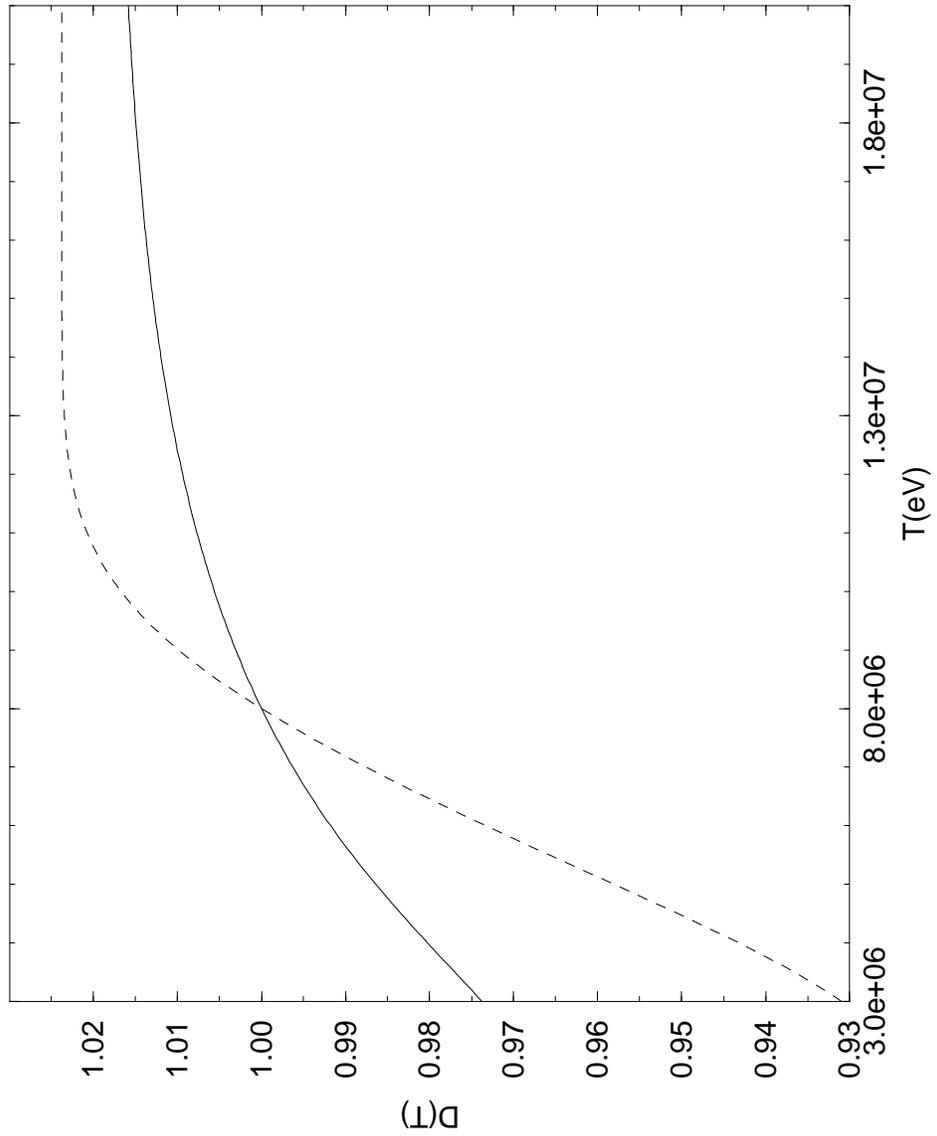,width=15cm}}
\end{picture}
\caption{Same as fig. 6 for the sixth field profile in table I. This is
essentially the same as the previous one except that it is multiplied by a 
scale factor of order 3 above the upper radiative zone. Notice that the
distortion is slightly increased with respect to fig. 6 indicating a
correlation between magnetic field intensity (i. e. solar activity) and distortion.}
\end{figure}


\end{document}